\documentclass[12pt]{spieman}  
\usepackage{amsmath,amsfonts,amssymb}
\usepackage{graphicx}
\usepackage{setspace}
\usepackage{tocloft}
\usepackage{soul}
\usepackage{comment}

\title{Image persistence flagging for SPHEREx}

\author[a,*]{Candice M. Fazar}
\author[c]{Charles D. Dowell}
\author[c]{Brendan P. Crill}
\author[b]{Phil Korngut}
\author[b]{Chi Nguyen}
\author[b]{Howard Hui}
\affil[a]{Rochester Institute of Technology, Center for Detectors, College of Science, 1 Lomb Memorial Drive, Rochester, NY, 14623}
\affil[b]{California Institute of Technology, 1200 E California Blvd,  Pasadena, CA 91125}
\affil[c]{Jet Propulsion Laboratory, California Institute of Technology, 4800 Oak Grove Drive, Pasadena, CA 91011}

\cftpagenumbersoff{figure}
\cftpagenumbersoff{table} 
\begin{document} 
\maketitle

\begin{abstract}
Image persistence in HAWAII-2RG HgCdTe detectors has been observed by multiple parties. Also known as latent signal, this effect occurs when sensor images following an illumination show a decayed form of the illuminated image even though the source has been removed and the detector has been reset. Using data from an engineering grade detector array delivered for SPHEREx testing illuminated with a wide range of fluxes, we demonstrate an interpretation and a working model from which the decaying signal can be estimated, providing the ability to flag pixels subject to excess persistence current beyond a specified threshold. Simulated persistence images allow validation of the module and prediction of its affect on flight data.
\end{abstract}


{\noindent \footnotesize\textbf{*}Candice M. Fazar,  \linkable{cmfsps@rit.edu} }

\section{Introduction}
\label{sect:intro}  

After exposure to light, many existing Teledyne HAWAII-2RG HgCdTe infrared detector arrays are subject to varying degrees of persistence current, which manifests as excess measured current in a decayed form of the original image.  This excess current still exists after array resets and power cycling, and slowly decays away with time.  Due to the potentially long time constants of this decay, subsequent images will be contaminated by this excess current thus interfering with accurate photocurrent measurements.  The SPHEREx mission, which employs six of these detector arrays, will carry out an all sky survey through a sequence of exposures separated by spacecraft slews, during which the detectors will experience frequent resets while exposed to sky light.\cite{2020Crill}  It is expected that after exposure to a bright source, the residual persistence current will contaminate subsequent sky images, even accounting for slew time and additional resets.  Such contamination is an important systematic error that must be mitigated for SPHEREx to achieve its science goals, particularly in characterizing the extragalactic background light.

Authors such as Smith\cite{2008Smith}, Tulloch\cite{2019Tulloch} and Goff\cite{2022Goff} describe the source of such behavior as charge traps primarily in the depletion region.  These traps collect charges during illumination and are not immediately swept away when the diode is reset.  Rather, they release at random times following the reset, resulting in a decaying current that adds to any photo current present in the device.  The wide range of time constants amongst the trap population collectively results in an inverse time-shaped current decay when large groupings of pixels are examined post-illumination, with individual pixel behavior depending upon the localized trap density and population.\cite{2014Anderson, 2016Mosby}  Persistence current has also been shown to be temperature dependent\cite{tulloch2018persistence} and highly dependent upon the characteristics of an individual array, with some detector arrays heavily affected while others showing no detectable effect.  
While Teledyne has recently verified production of new arrays with a specialized process that nearly eliminates persistence current,\cite{2024Beletic} missions using existing arrays will still need to account for the effect in their data analysis pipelines.

The SPHEREx science data analysis pipeline includes an image persistence flagging module to identify pixels subject to persistence current beyond a threshold using the measured currents in the previous image and flags representing the quality of that fit.  
To inform the module with limited persistence data from flight detector arrays,
we base our model upon data taken with an engineering grade 2.5~$\mu$m cutoff HgCdTe HAWAII-2RG test detector array for SPHEREx ({\ttfamily BBDet}), which is not one of the six arrays selected for flight.  
Presuming the functional form of the model is representative of flight arrays, limited additional data on flight detector arrays will inform model parameters.
Without sufficient data to inform a pixel-by-pixel model from which one can subtract out persistence current, we aim
to provide an upper limit to the persistence current per pixel per image 
based upon the aggregate data for the entire array.  From this result we will
flag pixels with predicted persistence current above the threshold. 
Ultimately, the threshold for tolerable persistence current is set by effects on the science goals of SPHEREx.
However, with SPHEREx-like integrations, we have measured the noise current on the order of $0.1~e^-/s$.  Therefore, for the purposes of this analysis and module development, we baseline $0.1~e^-/s$ as the persistence threshold with a secondary more ambitious level of $0.01~e^-/s$ on the order of magnitude of the dark current.  This threshold will be revisited during mission operations and science data analyses.

\section{Data and Modeling}
\label{sect:data}

\subsection{Illumination}
\label{subsect:illumination}

\begin{figure}[h]
\begin{center}
\begin{tabular}{c}
\includegraphics[height=6.5cm]{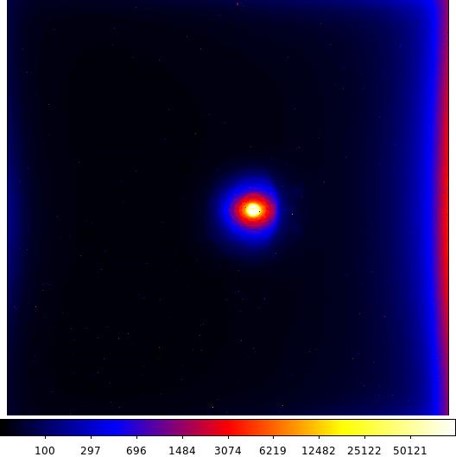}  
\end{tabular}
\end{center}
\caption 
{ \label{fig:illumination}
The broadband illumination pattern of detector array {\ttfamily BBDet} through a pin hole in an aluminum cover taped directly over the detector array's metal housing is shown with a logarithmic color scale in electrons per second.  The central bright spot is over 100,000 $e^-/s$ and significant light leakage around the cover is observed along one edge.} 
\end{figure}

Detector array {\ttfamily BBDet} was illuminated by a broadband source through a pin hole in an aluminum cover taped directly over the metal housing of the detector array.  This resulted in a concentrated illumination near the center of the array with a significant leakage around the cover along one edge where the cover was not completely light-tight.  The illuminating pattern can be seen in Fig.~\ref{fig:illumination}.

\begin{figure}[h]
\begin{center}
\begin{tabular}{c}
\includegraphics[height=5.1cm]{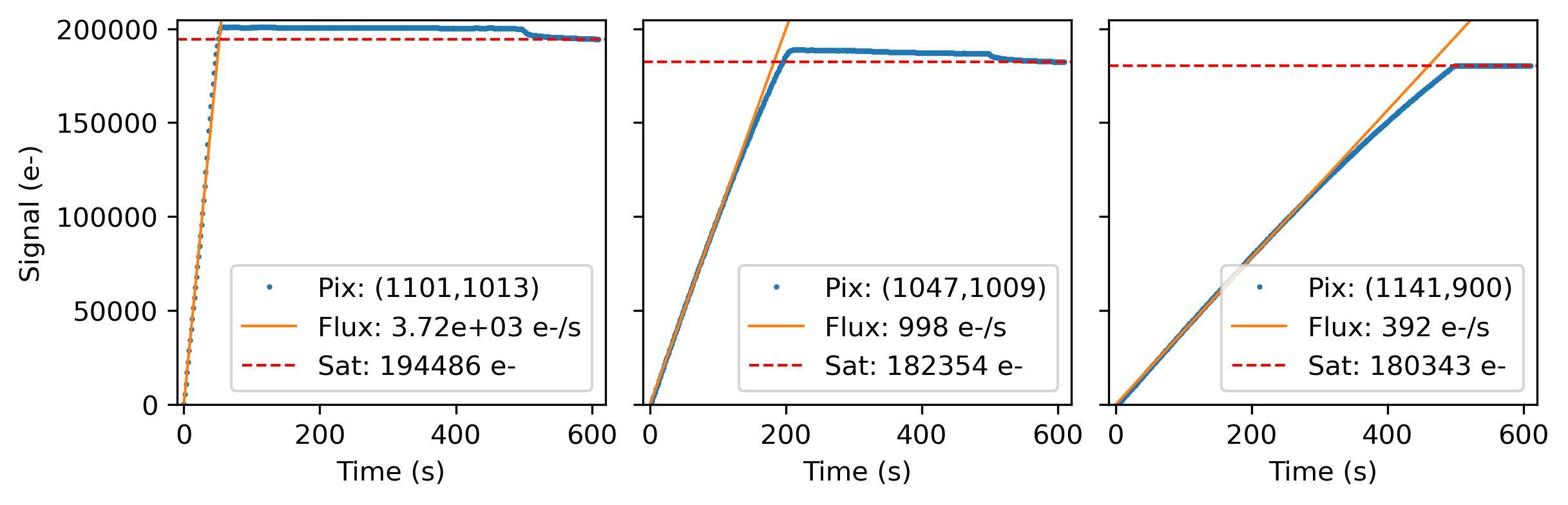}  \\
\hspace{1.0cm} (a) \hspace{4.0cm} (b) \hspace{4.0cm} (c) \\
\end{tabular}
\end{center}
\caption 
{ \label{fig:SUTRillumination}
SUTR plots for selected pixels. Data are plotted as blue dots, representing the collected signal as a function of time during the long illuminated integration.  The flux is determined based upon a line fit to the data up to 60\% of saturation, excluding the saturation region as can be seen in (b) and (c), resulting in the solid orange line.  
The shutter closes at approximately eight minutes into the exposure, causing pixels (a) and (b) to decay back down to a saturation level and pixel (c) to stop accumulating charge.  The saturation level is determined by the average of the last five samples of the signal level, illustrating the decay from forward bias for overly saturated pixels such as (a) and (b), and underestimating the well capacity for pixels such as (c).
 } 
\end{figure}

The data consists of a 613 second sample-up-the-ramp (SUTR) integration.  At 481 seconds ($\sim$8 minutes) into the integration, the source was powered off and the optical path blocked by a cold shutter.  The SUTR data and slope-fits for selected pixels are shown in Fig.~\ref{fig:SUTRillumination}.  These plots show various pixel behaviors and illustrate flux and saturation measurements for pixels subject to a range of illumination levels.
Heavily saturated pixels will saturate in forward bias and thus decay back down to the saturation level when the shutter is closed, as can be seen in panels (a) and (b) of Fig.~\ref{fig:SUTRillumination}.  Only data points below half of the saturation level at were used in determining the fit, thus avoiding saturation and minimizing nonlinearity.  Pixels with fewer than three data points below half-saturation were assigned zero flux.  
With a saturation level of approximately 200,000~$e^-$, the minimum flux necessary to saturate a pixel during this illumination is 416~$e^-/s$.

\subsection{Initial Persistence Current Response}
\label{subsect:initialpercur}
\begin{figure}[h]
\begin{center}
\begin{tabular}{c}
\includegraphics[height=5.1cm]{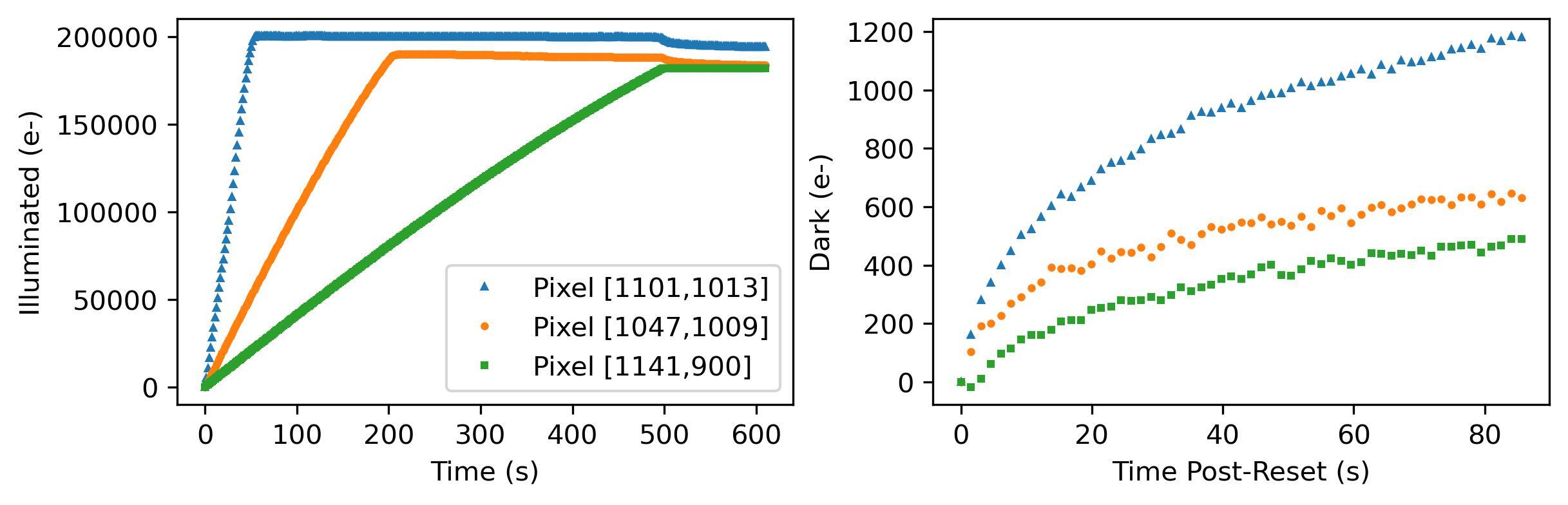}  \\
(a) \hspace{6.0cm} (b) \\
\end{tabular}
\end{center}
\caption 
{ \label{fig:SUTRpostillumination}
SUTR plots for selected pixels comparing (a) their illumination to (b) the accumulated persistence charge in the first dark integration post-illumination. The charge accumulation shown in panel (b) is strongest immediately following pixel reset and thus the downward curving non-linear trend of persistence signal is evident.  The heavily illuminated pixel in blue shows correspondingly significant persistence current.
} 
\end{figure}

Upon resetting the array after this long integration, the persistent signal begins to accumulate, manifesting as a non-linear, downward curving SUTR signal.  This is most heavily present in the SUTR immediately following illumination as shown in Fig.~\ref{fig:SUTRpostillumination}.  The downward curving trend of the persistence signal isn't expected to cause any flag conditions and thus SPHEREx on-board processing will perform a least-squares line-fit to the resulting data.
Therefore, these data, as well as the subsequent integrations taken every three minutes over the following fourteen hours were line-fit to simulate the on-board processing of the data.\cite{2016Zemcov}  

Individual pixels from the data in this initial post-illumination integration were analyzed in addition to groupings of pixels with approximately the same illuminating flux.  It was determined that the overall behavior of these pixels depended upon the extrapolated fluence, which is assumed equivalent to the accumulated charge for pixels far from saturation and extrapolated for saturated pixels by the multiplication of the measured flux and the illumination time.

\begin{figure}[b]
\begin{center}
\begin{tabular}{c}
\includegraphics[height=6cm]{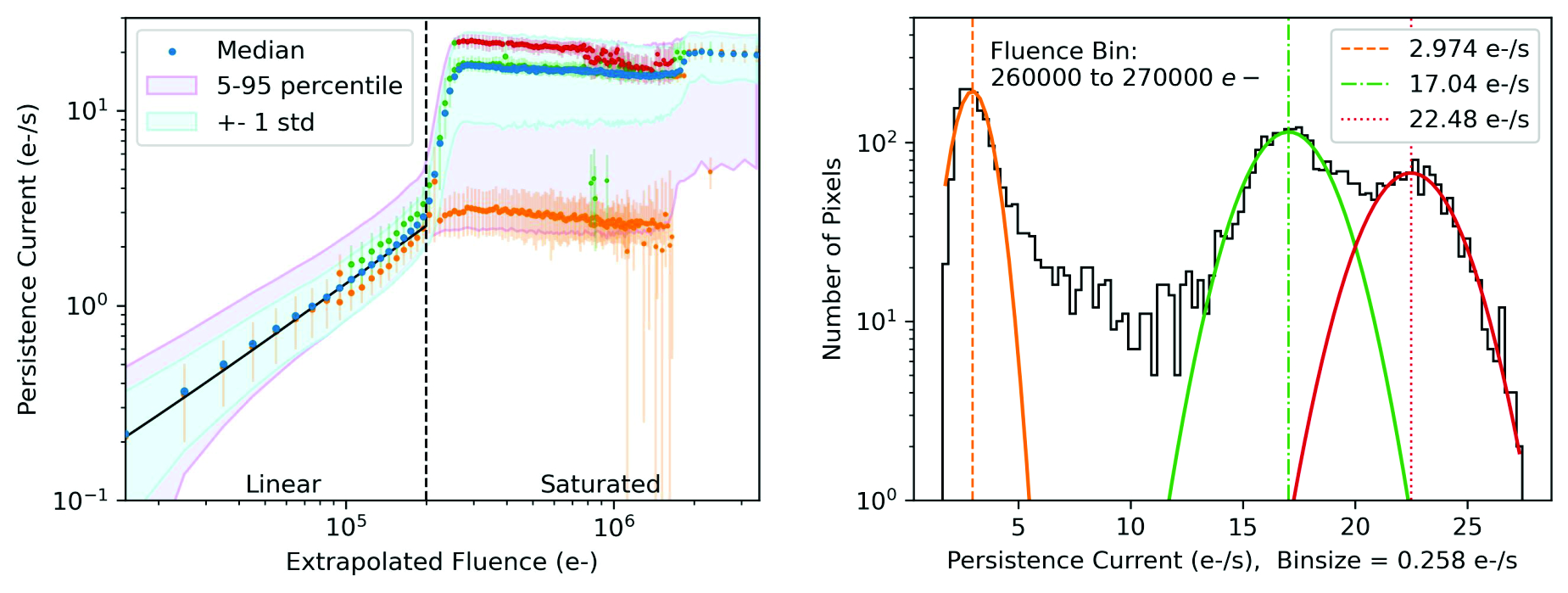}  \\
(a) \hspace{7.0cm} (b) \\
\end{tabular}
\end{center}
\caption 
{ \label{fig:percur_params_v_fluence_w_histobins}
Persistence current behavior.  
(a) Wide bins over the full range of illumination compare the approximately linear sigma-clipped median response illustrated by the overlapping solid black line in the non-saturated region to the abrupt change at saturation, which is shown by a vertical dashed line at 200,000 $e^-$.  In addition to the blue dots representing the median, the blue shaded region showing $\pm1$ standard deviation about the mean and the purple shaded region representing the 5th to 95th percentile response, orange, green and red dots represent the detected Gaussian peaks from low to high persistence currents respectively with error bars representing the $\pm \sigma$ of the Gaussian fit such as those shown in panel (b).  These illustrate the range of persistence current response to the same illumination level.
In addition, the size of the point is proportional to the magnitude of the number of pixels in the bin or the histogram peak, showing relative populations.
Noticeably, the peak around 3.0~$e^-/s$ decreases in population relative to the upper peaks, resulting in the increase in the lower end of the percentile range.
(b) Histogram showing 
the persistence currents displayed by pixels within the extrapolated fluence bin from 260000 to 270000 $e^-$.
The three peaks indicated by this histogram at 3.0, 17 and 22.5 $e^-/s$, are typical of all bins after the steep rise at the onset of saturation, as can be seen in panel (a).  
} 
\end{figure}

Blue dots in panel (a) of Fig.~\ref{fig:percur_params_v_fluence_w_histobins}
show the sigma clipped median line-fit current observed in that first integration after the reset following illumination for bins of pixels with similar extrapolated fluence. Below saturation at approximately 200,000 $e^-$, illuminated pixels exhibit a low-level persistence current that 
is approximately linearly dependent upon extrapolated fluence.
Recent measurements with different illumination times have verified this by showing that the illumination time is directly proportional to the persistence current within a given bin of a non-saturating flux. This produces the same predicted persistence current as shifting to the corresponding non-saturated fluence bin.  Since SPHEREx's observation plan involves integrations of equal duration, 
measured flux will be linked to a specific extrapolated fluence bin throughout the duration of flight.

At saturation the median response increases dramatically due to many pixels in each saturating fluence bin having significantly higher persistence current.  There is some evidence to suggest that blooming\cite{2021Blooming} may be a contributing factor to this rise, but further investigation is necessary to confirm.  Equally significant is that some pixels do not display this dramatic increase in persistence current, but rather maintain low persistence currents even under heavy illumination.  To characterize this, a histogram of each fluence bin was plotted to show the range of persistence currents resulting from the same measured illumination.  One such bin is shown in panel (b) of Fig.~\ref{fig:percur_params_v_fluence_w_histobins}.  
For pixels not subject to blooming, different responses to the same illumination likely result from different trapping populations.  Without the data necessary to distinguish between these pixel populations, we model against the sigma-clipped median response and flag according to the maximum possible persistence current per bin, keeping in mind that our primary goal is to flag pixels contaminated by persistence current at any level above the threshold.

\subsection{Decay Timing}
\label{subsect:timing}

\begin{figure}[h]
\begin{center}
\begin{tabular}{c}
\includegraphics[height=7.0cm]{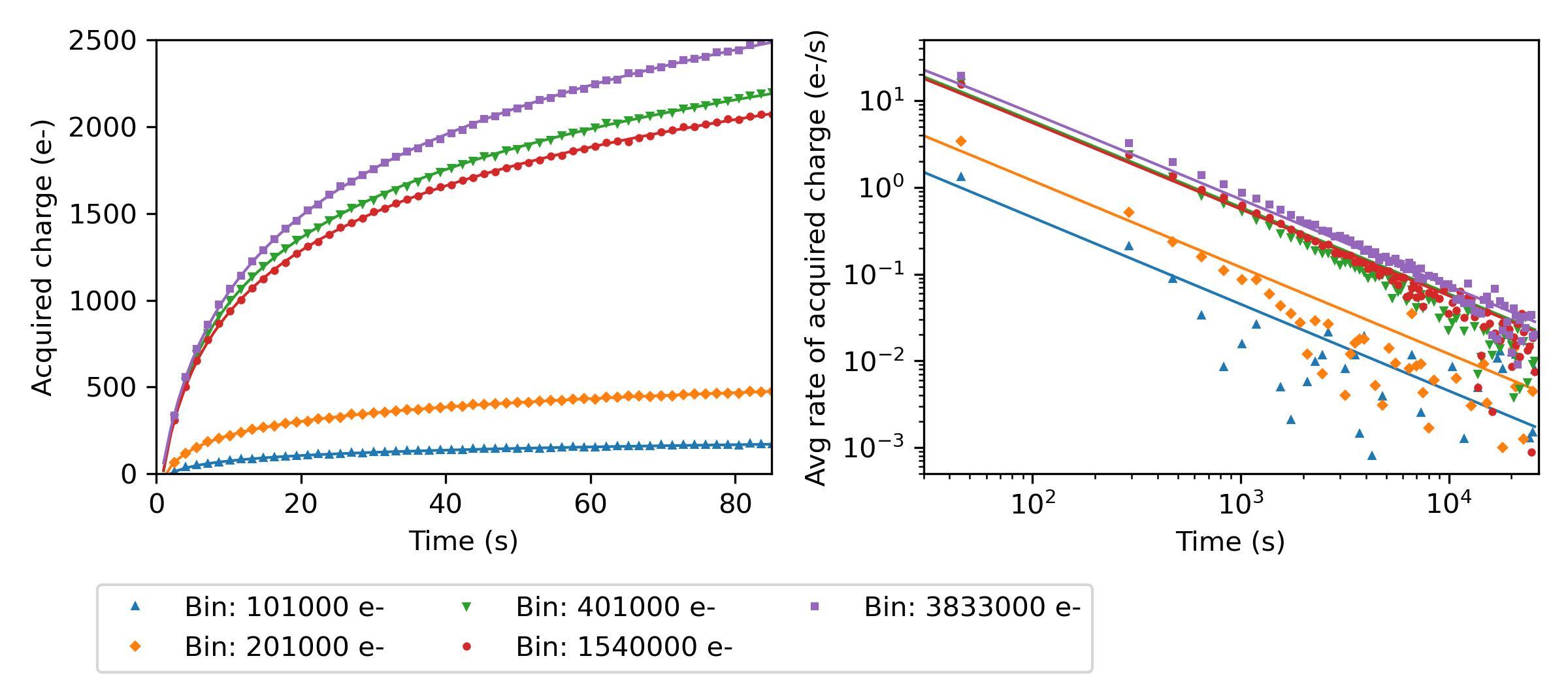}  \\
\end{tabular}
\end{center}
\caption 
{ \label{fig:percurtiming}
Persistence current modeling and timing for selected fluence bins.  {\it Left.} The accumulation of persistent charge versus time in the initial integration following illumination are plotted with their overlapping logarithmic fits.  {\it Right.} A temporal derivative of the logarithmic fit from the left panel is plotted as a solid line to predict the time dependence of the persistence current.  Symbols represent the slope of a linear fit to each sequential dark integration following illumination placed at the midpoint of the measurement. Fit parameters are tabulated in Tab.~\ref{tab:modelparameters}.}
\end{figure}

Persistence charge accumulation in the first integration following illumination has been characterized for the sigma-clipped median 
accumulated signal for each bin.
As shown in the left panel of Fig.~\ref{fig:percurtiming},
selected bins are fit to 
\begin{equation}
\label{eq:logfit}
q(t) = A_0 \ ln(t + \tau) + q_0,
\end{equation}
where $t$ represents the time after the reset and $A_0$, $\tau$, and $q_0$ are parameters to be fit representing persistence charge, offset time and initial charge respectively.  These parameters are then used to extrapolate the persistence current as a function of time by the derivative of Eq.~\ref{eq:logfit}, 
\begin{equation}
\label{eq:invT}
i(t) = \frac{A_0}{t + \tau}.
\end{equation}
These predictions are then compared with the sigma clipped median line-fits of the same bins from  
all exposures over the fourteen-hour-long set of dark measurements as shown in the right panel of Figure~\ref{fig:percurtiming}, with parameters given in Tab.~\ref{tab:modelparameters}.

\begin{table}[hb]
    \begin{center}
    \begin{tabular}{cccccc}
        Fluence & $A_0$ & $\tau$ & $q_0$ & Average Slope & $A_{bin}$ \\
        \hline
        101,000 $e^-$ & 45.1 $e^-$ & -0.01 $s$ & -30.9 $e^-$ & 1.4 $e^-/s$ & 54.1 $e^-$\\
        201,000 $e^-$ & 120.2 $e^-$ & 0.32 $s$ & -60.8 $e^-$ & 3.6 $e^-/s$ & 144.2 $e^-$\\
        401,000 $e^-$ & 592.6 $e^-$ & 1.19 $s$ & -449.4 $e^-$ & 17.2 $e^-/s$ & 711.1 $e^-$\\
        1,540,000 $e^-$ & 564.1 $e^-$ & 1.25 $s$ & -438.4 $e^-$ & 16.3 $e^-/s$ & 676.9 $e^-$\\
        3,833,000 $e^-$ & 731.7 $e^-$ & 2.16 $s$ & -781.5 $e^-$ & 20.3 $e^-/s$ & 878.0 $e^-$\\
        \\
    \end{tabular}
    \caption{Parameters for the fitted logarithmic model of  Eq.~\ref{eq:logfit} for selected bins plotted in the left panel of Fig.~\ref{fig:percurtiming} and the average slope resulting from the linear fit to the same data, which results in the first data point for each bin in the right panel of the same figure.  $A_{bin}$ is $A_0$ amplified by 1.2.
    \label{tab:modelparameters}}
    \end{center}
\end{table}

As can be seen in the right panel of Fig.~\ref{fig:percurtiming}, the model underestimates the current obtained by a line-fit to the first data set, from which the parameters come.  It is therefore plausible that the underestimation is in part due to line fitting a non-linear behavior, which most heavily impacts the first few integrations after exposure to a bright source.  To compensate for this we amplify our fit by a factor of 1.2 resulting in $A_{bin}$ in Tab.~\ref{tab:modelparameters}.
This allows our predictions to better 
estimate the measured persistence current.  We find that with the amplification for these data, the sigma-clipped median follows or drops below the inverse time functional dependence at later times.

\begin{figure}[ht]
\begin{center}
\begin{tabular}{c}
\includegraphics[height=6.2cm]{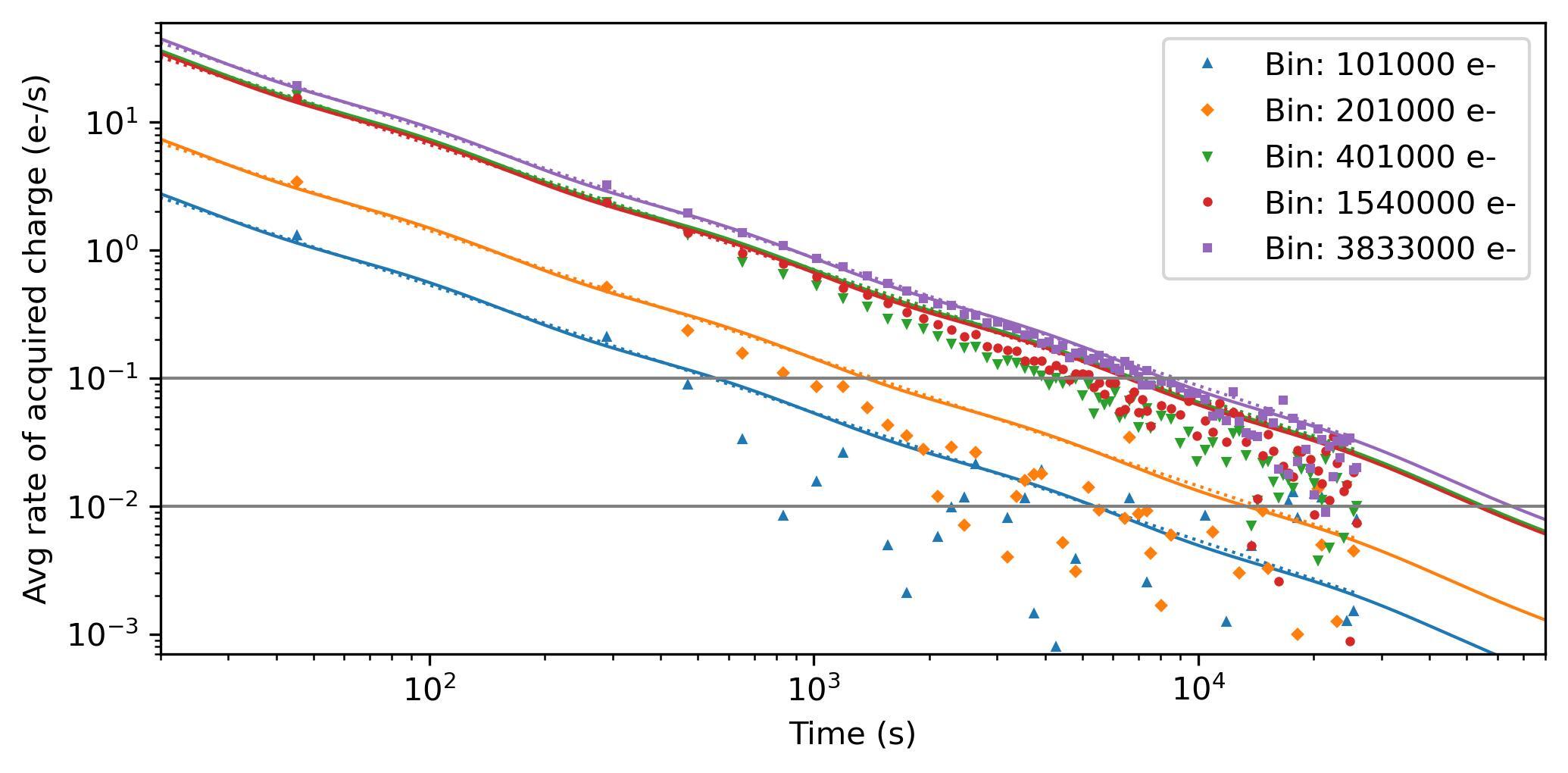}  \\
\end{tabular}
\end{center}
\caption 
{ \label{fig:exp_decay}
Persistence current as a function of time with model comparison.  Symbols represent the slope of a linear fit to sequential dark integrations, same as Fig.~\ref{fig:percurtiming}.  Models are amplified by a factor of 1.2 to account for underfitting of the logarithmic model.  The amplified exponential decay summation model given by Eq.~\ref{eq:bbdetexpsum} is plotted as a solid line and the amplified inverse time model with $\tau$ set to one is plotted as a dashed line in the same color as the symbols.  Two threshold possibilities are shown as horizontal grey lines, the higher corresponding to the order of magnitude of noise in the current measurement and the lower corresponding to the order of magnitude of dark current.}
\end{figure}

Since our ultimate goal is to flag pixels subject to persistence current as the flux varies from exposure to exposure, approximating the inverse time behavior with a sum of exponentials is a computationally efficient method that retains information about the history of exposures that preceded it, allowing one to continue the predictive model from any exposure in the sequence.
Figure~\ref{fig:exp_decay} shows this model for selected bins.  
In this approximation, the threshold above which persistence current is flagged determines the necessary number of terms in the summation.  Note that the number of coefficients and the spacing between them affect the quality of the model.  
For SPHEREx we have opted to use six terms, which is sufficient to handle the large initial persistence currents for pixels subject to high fluence that result in long durations of persistence current above the lower dark current threshold of $0.01~e^-/s$, while keeping the undulations of the model to a minimum.  This model, illustrated in Fig.~\ref{fig:exp_decay}, is given by Eq.~\ref{eq:bbdetexpsum},
\begin{equation}
\begin{aligned}
\label{eq:bbdetexpsum}
i_p(t) = A_{bin} \times (0.183e^{-0.1t}+0.029e^{-0.0158t}+0.0046e^{-0.00251t}+0.000729e^{-0.000398t} \\ 
+ 0.000115e^{-6.31\cdot 10^{-5}t}+1.83\cdot 10^{-5}e^{-10^{-5}t}),
\end{aligned}
\end{equation}
with the bin-dependent initial current given by $A_{bin}$, which is proportional to $A_0$ from the logarithmic fit. The remaining summation is normalized so that it most closely approximates the functional dependence of Eq.~\ref{eq:invT} with $A_0=1$ and $\tau=1$ s and is therefore bin-independent. We use this model as a starting point for the flagging module.

\section{Persistence Current Flagging Module}
\label{sect:flagging}

With this model 
we flag pixels according to the following general steps.

\begin{enumerate}
\item Determine photo current in exposure.
\item Extrapolate fluence to find behavior range.
\item Determine model amplitudes.
\item Apply model to get an estimate of persistence current from exposure.
\item Add estimate to decaying value of persistence current due to previous exposures.
\item Flag pixels with persistence current beyond specified threshold in subsequent exposure.
\item Repeat for each exposure.
\end{enumerate}

The most general form of the mathematical equation describing our model of persistence current for a single pixel as a function of $t$, the time after the reset at the end of the $k^{th}$ exposure, and allowing for ease of tracking persistence over multiple illuminations is given by
\begin{equation}
\label{eq:percurmodel}
i_p(t) = \sum_{n=1}^{N} A_n C_n e^{-t/\tau_n},
\end{equation}
where 
$N$ is the number of terms in the equation, $C_n$ is the normalization coefficient and $\tau_n$ is the time constant of the $n^{th}$ term respectively.
The amplitude $A_n$ is specified by
\begin{equation}
\label{eq:A_n}
A_n = A_n' e^{\frac{-t_{elapsed}}{\tau_n}} + A_k,
\end{equation}
where $A_n'$ is $A_n$ from the previous exposure, $t_{elapsed}$ is the time between exposures and $A_k$ is the persistence current response
corresponding to the extrapolated fluence in the $k^{th}$ exposure.  Following the $k^{th}$ exposure, the $N$ values of $A_n$ are updated per pixel and the summation of Eq.~\ref{eq:percurmodel} is then evaluated, providing an estimate of persistence current present per pixel in exposure $k+1$.  This persistence ``image'' is then saved for evaluation purposes and pixels with predicted currents above the threshold are flagged.

\subsection{Early Release}
\label{sect:earlyrelease}
In lieu of sufficient data to create a pixel by pixel model, we overestimate by predicting to the greatest possible persistence current under the given illumination conditions.  We use measured currents to predict extrapolated fluence and estimate the worst case response. 
However, treating all pixels equally has its drawbacks as not every pixel has the trapping conditions necessary to produce the maximum expected persistence current.  Consequently, to ensure that the pixels are not flagged for longer than necessary, the pixel's measured current in a given exposure will be compared to the predicted persistence current in that same exposure.  If the persistence current model predicts that the pixel should be flagged, but it is determined that the measured current is below an early-release threshold for three subsequent exposures, it will be determined that the pixel's persistence current has prematurely subsided and the pixel will then be released from the flag and the persistence current model reset to zero.

\subsection{Transients}
\label{sect:transients}
Cosmic rays that pass through the detector will produce a sudden increase in charge in the SUTR data.  On-board processing for SPHEREx will detect jumps of a sufficient level to deviate slope measurements and flag these pixels for transients.  In such a case, since on-board processing will remove information relevant to the magnitude and timing of the jump, the pixel will also be flagged for persistence and the model will use the maximum possible amplitude in its calculation for the subsequent exposure.  Such pixels will be likely to have over-predicted persistence currents and therefore be released from the flag by the early release check.

\section{Model Parameters and Validation}

\begin{figure}[ht]
\begin{center}
\begin{tabular}{c}
\includegraphics[height=6.2cm]{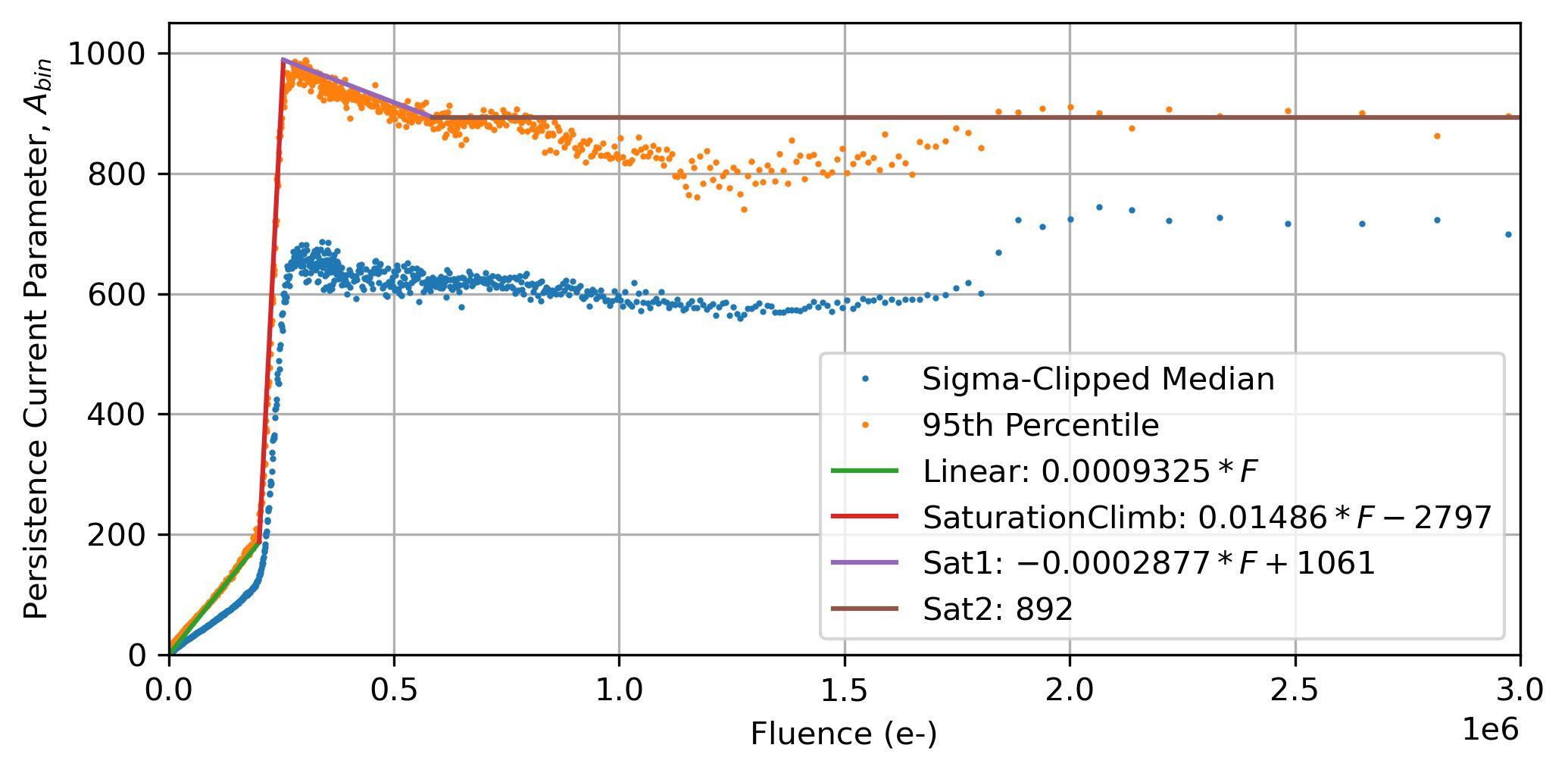}  \\
\end{tabular}
\end{center}
\caption 
{ \label{fig:modelAparams}
Persistence current parameter $A_{bin}$ as a function of fluence. Sigma-clipped median response (blue) and higher 95th percentile response (orange) give parameters for $A_{bin}$ by letting $\tau=1$ in Eq.~\ref{eq:invT} 
and scaling the measured current to $t=0$.
The functional fit which provides $A_k$ as a function of fluence was implemented on the 95th percentile in order to over-estimate the persistence response for the majority of pixels. 
} 
\end{figure} 

Using the data taken with {\ttfamily BBDet}, the 
sigma-clipped median average persistence current and the 95th percentile average current of the first thirty seconds of the first dark exposure following illumination for each bin was used to determine $A_{bin}$ by letting $\tau=1$ in Eq.~\ref{eq:invT} 
and scaling the measured current to $t=0$.
The resulting parameters for the sigma-clipped median and the 95th percentile are shown in Fig.~\ref{fig:modelAparams}. 
Using the sigma-clipped median values to estimate the persistence response would underestimate the persistence current for approximately half of the array pixels.  Therefore, to ensure overestimation for conservative flagging we fit the 95th percentile response to a piece-wise linear function to get the parameter $A_{k}$ in Eq.~\ref{eq:A_n} as a function of illuminating fluence.

\begin{figure}[h]
\begin{center}
\begin{tabular}{c}
\includegraphics[height=6.2cm]{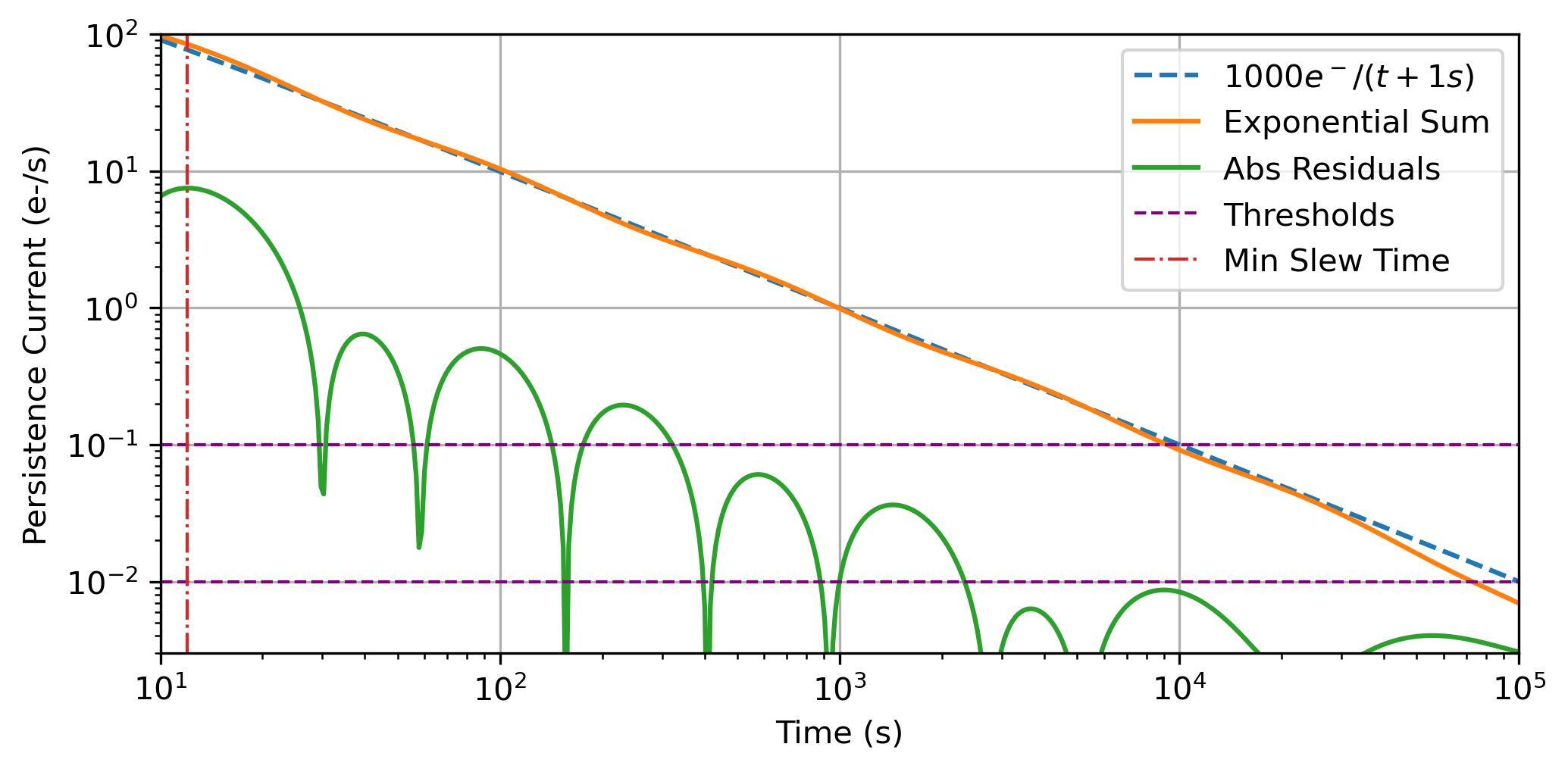}  \\
\end{tabular}
\end{center}
\caption 
{ \label{fig:residuals}
 Residuals between the exponential and inverse time models for the amplified $A_{bin} = 1000$ and chosen $C_n$ coefficients shown in Eq.~\ref{eq:bbdetexpsum}.  This model includes sufficient time constants to closely approximate the expected behavior from the minimum slew time through the time at which the current approaches the lower flagging threshold of $0.01~e^-/s$ while keeping the undulations of the model to a minimum.  
 }
\end{figure}

The parameters $C_n$ are chosen so that Eq.~\ref{eq:percurmodel} most closely approximates Eq.~\ref{eq:invT} from the minimum calculation time through the maximum decay time necessary to reach the threshold limits.  The resulting summation meeting these conditions with six terms is shown in Eq.~\ref{eq:bbdetexpsum}.  These two models are compared in Fig.~\ref{fig:residuals}, showing the exponential summation model approximating
the inverse time decaying persistence current from the minimum slew time through the time at which the current crosses the flagging threshold. 
In this figure, the absolute difference between the two models remains an order of magnitude below the predicted currents past the $0.1~e^-/s$ threshold.

\begin{figure}[h]
\begin{center}
\begin{tabular}{c}
\includegraphics[height=12.0cm]{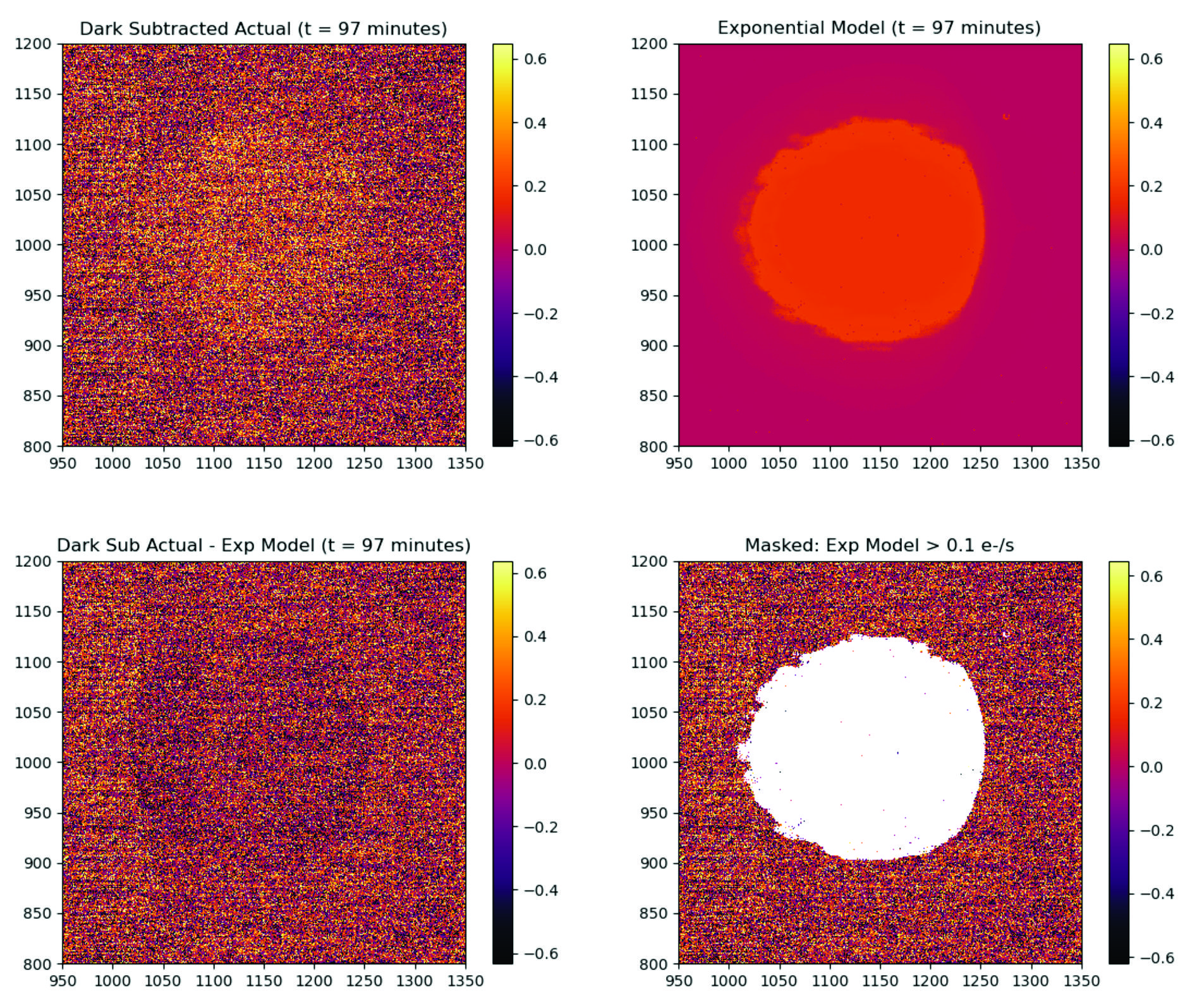}  \\
\end{tabular}
\end{center}
\caption 
{ \label{fig:permodelimages}
Actual persistence images compared to predictions at $t = 97$~minutes into the dark sequence. 
Top-Left: The dark-subtracted persistence image produced from slope-fits to frames 1-20 of the SUTR data taken 97 minutes after the first reset of the detector array following illumination.  Top-Right: The exponential model prediction on the same scale as the dark-subtracted image.  Bottom-Left: Difference image showing the exponential model subtracted from the top-left image. This image shows that the model over-estimates the persistent signal in this image.  Bottom-Right: Dark-subtracted persistence image with flagged pixels masked (white).}
\end{figure}

This model was then verified by generating images predicting the data as shown in Figure~\ref{fig:permodelimages}.  These panels illustrate the over-estimation of the model and the flagged result of affected pixels at 97 minutes into the dark sequence after the saturating illumination.  At this time, all non-saturated pixels should be decayed below the threshold, leaving only the most strongly saturated pixels still on the ramp.
It was found during this analysis that the shortest time constants had a strong boost, particularly for pixels with fluence between $200000$ and $500000~e^-$.  It is uncertain why this is the case, but there is some evidence to suggest that blooming may again be a contributing factor.  This situation will be investigated in future analyses.  In addition, it is noted that groupings can be made of pixels across the array that exhibit similar persistence current response, likely a result of similar trap distributions.  An analysis of persistence in flight data would benefit from distinguishing between these populations.

\section{Simulations}
\begin{figure}[h]
\begin{center}
\begin{tabular}{c}
\includegraphics[height=2.5cm]{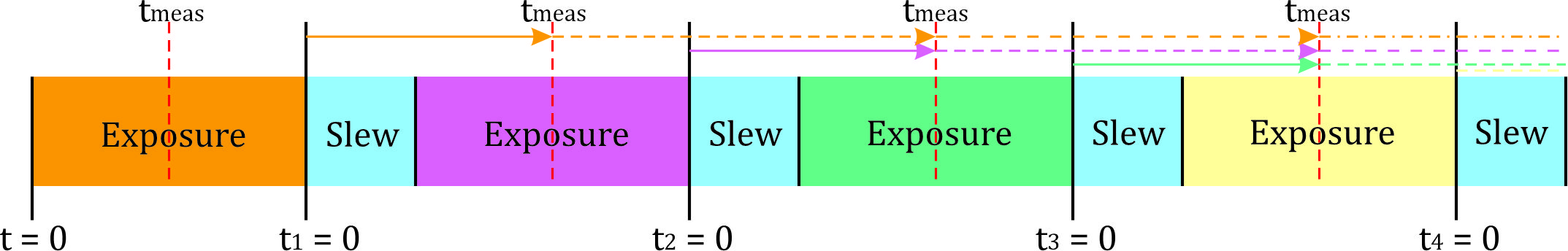}  \\
\end{tabular}
\end{center}
\caption 
{ \label{fig:timing}
Timing of persistence module.  Vertical black lines correspond to detector resets, with the resets following an exposure starting the ``persistence clock.''  It is at these resets that $A_0$ and subsequently $A_n$ are determined.  The horizontal lines along the top of this figure illustrate the onset and subsequent decay of persistence on sequential exposures. 
 Persistence calculation is evaluated mid-exposure at $t_{meas}$ with the previously determined $A_n$.  During the slew, frequent detector resets are designed to keep pixels from saturating, thus minimizing the effect of slewing on persistence.} 
\end{figure}

The next step in validating the module and testing flagging effectiveness is to simulate it with the SPHEREx Sky Simulator,\cite{2020Crill} which uses a sky model that includes compact and diffuse sources in addition to known detector array properties to produce simulated sky images.  By incorporating the persistence model into the simulator, we can simulate the percentage of pixels affected by persistence current as well as verify the module's performance with expected data.  
Such a task 
requires accurate knowledge of the model's timing, particularly because the persistence current is constantly being replenished as it decays, as illustrated by the colored horizontal lines along the top of Fig.~\ref{fig:timing}.  Persistence current resulting from the release of charges that are trapped during each exposure is initiated in the model at the first reset following that same exposure, indicated by a vertical black line.  
Given that the heaviest persistence current from any given exposure exists during the slew from one target to the next, we expect that frequent resets during slew will not only minimize persistence current due to bright targets present in the field of view at this time, it will also flush the strongest of the residual persistence current, resulting in lower persistence currents during the subsequent exposure. 
As a result, the flagging module ignores any sources present between fields and only considers data collected during each target exposure.  At each concluding reset $A_0$ is determined from the flux measurements in the completed exposure and $A_n$ is updated.  The summation of Eq.~\ref{eq:percurmodel} is then evaluated mid-exposure at the vertical red dashed lines, using $t=t_{meas}=t_{slew}+0.5t_{exposure}$.
This provides an estimate of the persistence current in each exposure.

\begin{figure}[h]
\begin{center}
\begin{tabular}{c}
\includegraphics[height=7.0cm]{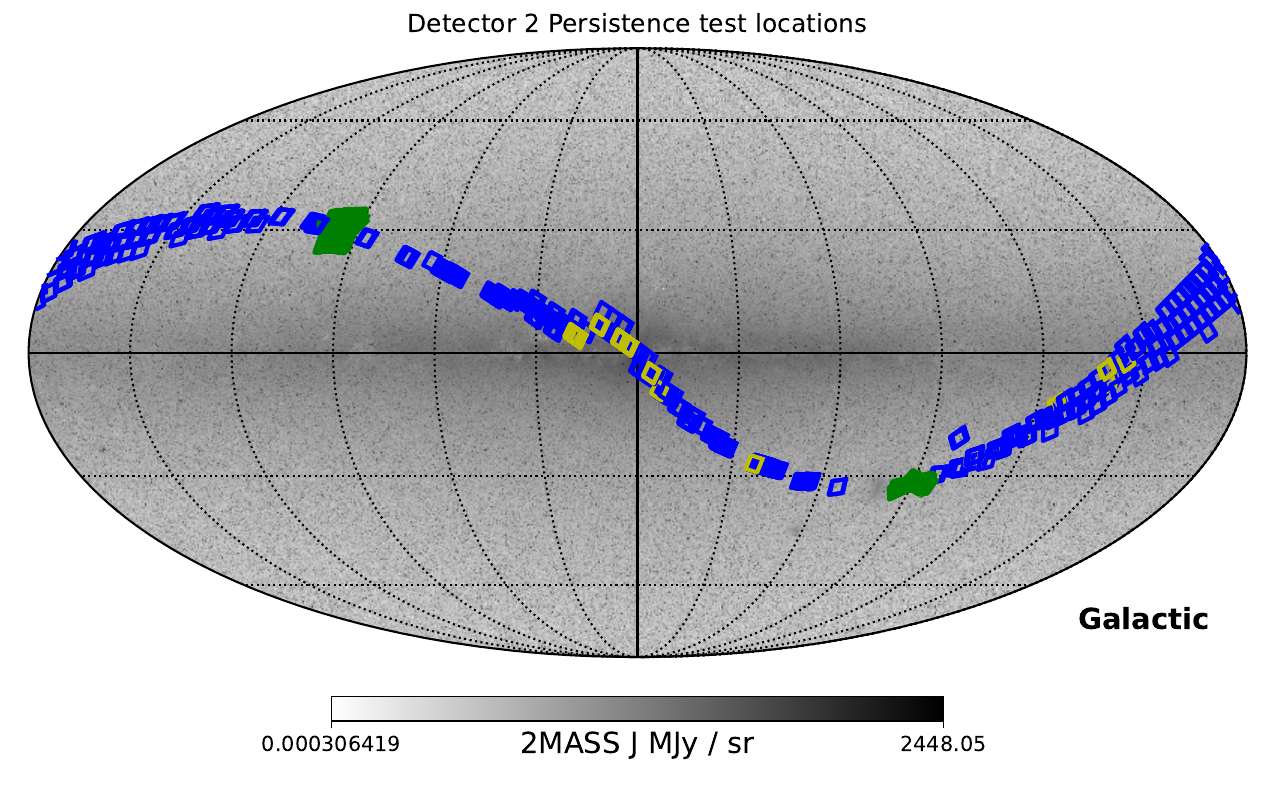}  \\
\end{tabular}
\end{center}
\caption 
{ \label{fig:fieldlocations}
Simulated field locations shown by blue squares with green regions indicating deep field observations and yellow squares indicating exposures taken in the South Atlantic Anomaly.} 
\end{figure}

\begin{figure}[h]
\begin{center}
\begin{tabular}{c}
\includegraphics[height=6.2cm]{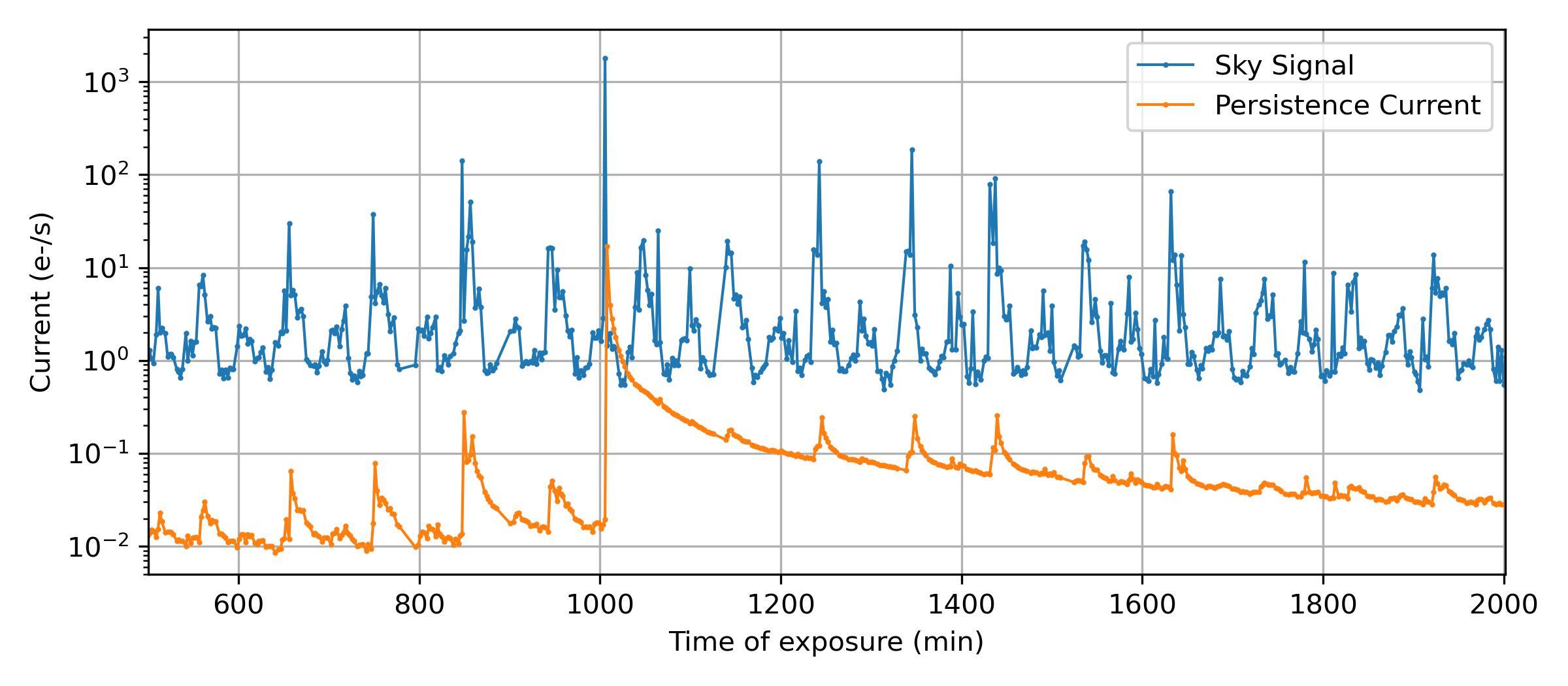}  \\
\end{tabular}
\end{center}
\caption 
{ \label{fig:singlepixsim}
Simulated single pixel persistence as a function of time over the simulated field locations in Fig.~\ref{fig:fieldlocations}.  A heavily saturating source is observed at 1000 seconds leading to a long decay time.} 
\end{figure}

\begin{figure}[h]
\begin{center}
\begin{tabular}{c}
\includegraphics[height=7cm]{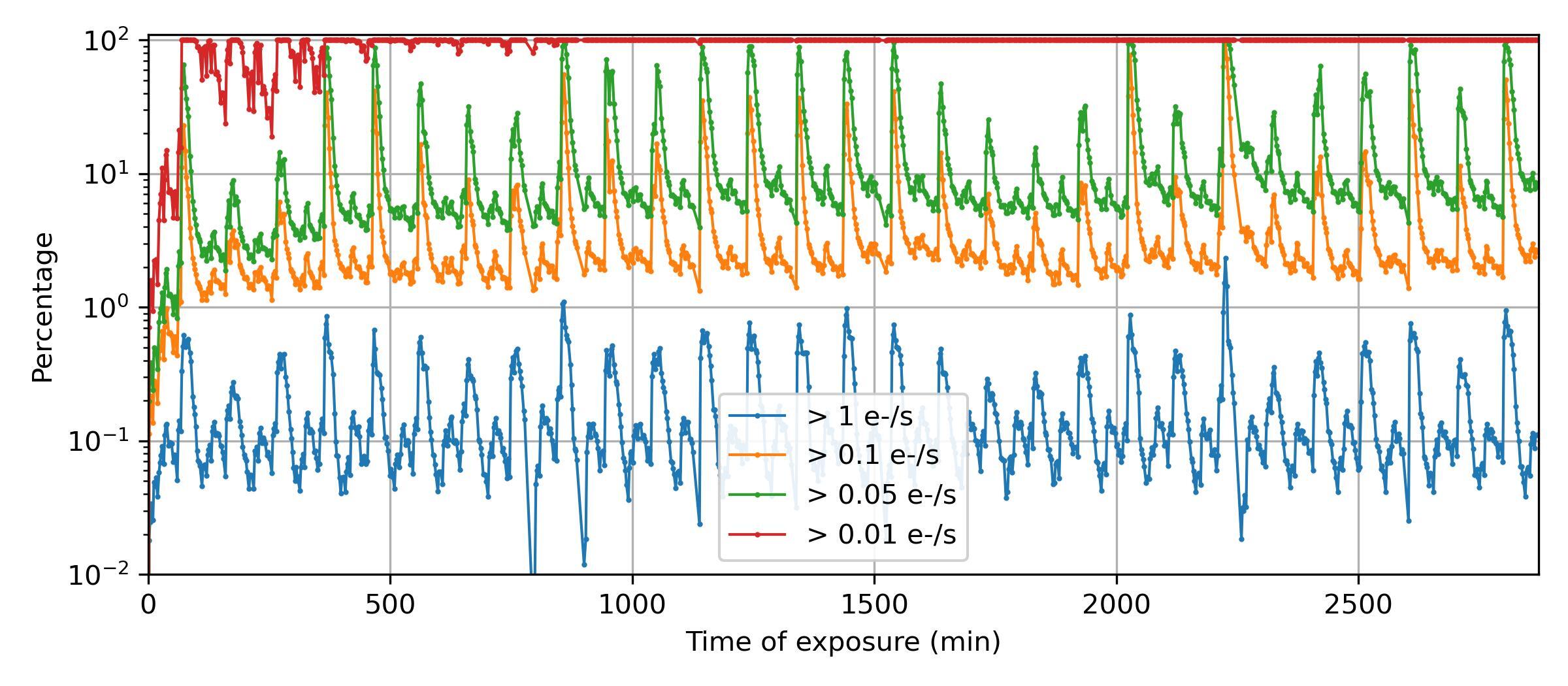}  \\
\end{tabular}
\end{center}
\caption 
{ \label{fig:simulations}
Simulated percentage of flagged pixels versus time over the simulated field locations in Fig.~\ref{fig:fieldlocations}. 
In this simulated view the long time constants following saturating illumination prevent pixels from decaying below 0.01~$e^-/s$ before another saturating source is observed.} 
\end{figure}

This model was implemented with two days of simulated sky images comprising 1143 exposures which include observations close to the Galactic center and Ecliptic poles as shown in Fig.~\ref{fig:fieldlocations}. 
In this sequence there are very highly illuminated exposures followed by minimally illuminated exposures, which highlights the impact of persistence on these situations.  To illustrate, corresponding persistence images were generated and the behavior of a single pixel is illustrated in Fig.~\ref{fig:singlepixsim} while the percentage of pixels flagged versus time is shown in Fig.~\ref{fig:simulations}. 

In the single pixel simulation the persistence current climbs quickly in the initial exposures and hovers above a persistence ``floor'' around $0.01~e^-/s$, on the order of magnitude of the dark current.  Right around 1000 minutes into the exposure, the pixel simulated in Fig.~\ref{fig:singlepixsim} is saturated  
by a significant source.
It then experiences a rapid decay to below $1~e^-/s$, but long time constants cause the persistent signal to remain above the threshold of $0.1~e^-/s$ for the following 200 minutes, affecting nearly 100 images.  Over the remainder of the simulated images, this pixel is predicted to stay above the lower threshold of $0.01~e^-/s$.  This illustrates the impact of heavily saturating flux on individual pixels.

 With the persistence model from array {\ttfamily BBDet},  the overall impact of continual observations including saturating sources and perpetual Zodiacal light contributions can be seen in Fig.~\ref{fig:simulations}.  It is expected that sources brighter than a magnitude of 12 will saturate pixels in at least one array and Zodiacal light will contribute a minimum of $0.002~e^-/s$ of persistence current to every observation. Fig.~\ref{fig:simulations} predicts that $100\%$ of pixels will experience a persistence ``floor'' above $0.01~e^-/s$, at least for the duration and pointings of this simulation and one can expect a minimum of $2\%$ of pixels to be flagged above the threshold of $0.1~e^-/s$ at all times.  
 This is suggestive that persistence current may be a limiting component of the measured signal, especially for low background pointings.

\section{Summary and Next Steps}

In summary, the goal of this work was to develop a model for persistence current that could be used to flag pixels potentially contaminated by persistence current above a specified threshold.  For this analysis we considered $0.1~e^-/s$ and $0.01~e^-/s$, representing the noise in the current and the dark current magnitude respectively.  To this end, we used persistence measurements of the test detector array {\ttfamily BBDet} including a long 10 minute integration including 8 minutes of illumination by a source with flux ranging from tens to hundreds of thousands of electrons per second followed by eight hours of dark measurements every three minutes to develop a model characterizing the persistence current in this detector array.  The persistence time-decay was found to be well represented by a function of inverse time, while the initial response was a function of extrapolated fluence.  This particular detector array demonstrated a linear dependence upon fluence in the non-saturated regime with multiple levels of persistence current demonstrated past saturation.  These different levels were attributed to different trapping populations, with blooming suggested as a potential contributing factor. To ensure pixels with persistence current beyond a threshold were flagged, the flagging module was described, implemented and demonstrated with parameters representing the 95th percentile of persistence response.  

It is expected that this model will overestimate the persistence current, allowing us to flag pixels as potentially contaminated with persistence current beyond the specified threshold.  The early release module was not tested in this analysis, but is expected to mitigate overestimating the persistence current in flight data.
Presuming this model is representative of the flight arrays, persistence images were produced based upon images generated by the SPHEREx Sky Simulator and the effects of persistence upon two days of simulated data were analyzed and found to reach a non-zero steady-state behavior, indicating that persistence current may be a limiting component of the measured signal.
This simulation also illustrated the percentage of pixels flagged based upon the chosen threshold level, which will help the science team to analyze the impact of persistence on the science data.  For the noise threshold of $0.1~e^-/s$ we observed approximately 2\% of pixels subject to persistence at all times. For science investigations requiring persistence currents below this level, persistence may be a significant issue.  Thus, the threshold will be revisited during mission operations and science data analysis.

The next steps of processing and interpreting persistence data on flight arrays are already underway and expected to result in updated model parameters.  With the new flight-array specific parameters, similar evaluations will be performed to understand the effects of persistence on flight data.  After SPHEREx launches in early 2025, in-flight characterization of persistence effects will be a challenge due to lack of a shutter in the SPHEREx optics.  Verification of the effectiveness of the flagging technique will be evaluated in a future investigation using cross correlations of sequential exposures.

\clearpage

\section*{End Pages}

\subsection*{Disclosures}
This author declares no conflicts of interest in publishing this information.

\subsection* {Code and Data Availability} 
Code and data may be obtained upon request.  Contact the lead author at \linkable{cmfsps@rit.edu}.

\subsection* {Acknowledgments}
We acknowledge support from the SPHEREx project under a contract from the NASA/GODDARD Space Flight Center to the California Institute of Technology.

Part of the research described in this paper was carried out at the Jet Propulsion Laboratory, California Institute of Technology, under a contract with the National Aeronautics and Space Administration (80NM0018D0004).


\bibliography{report}   
\bibliographystyle{spiejour}   

\end{document}